\def\gsim{~\rlap{$>$}{\lower 1.0ex\hbox{$\sim$}}}
\def\lsim{~\rlap{$<$}{\lower 1.0ex\hbox{$\sim$}}}

\def\h2o{\rm{H_{2}O}}
\def\mh2{\rm{H_{2}}}

\def\co2{\rm{CO_{2}}}
\def\ch4{\rm{CH_{4}}}
\def\nh3{\rm{NH_{3}}}
\def\n2{\rm{N_{2}}}

\documentclass[12pt, preprint]{aastex}
\usepackage{graphicx}
\usepackage{natbib,color,lscape}
\usepackage{subfigure}
\usepackage{threeparttable}
\usepackage{longtable}
\usepackage{url}
\usepackage{subfigure}
\usepackage{rotating}

\begin{document}
\title{Habitable Zone Boundaries for Circumbinary Planets}
\author{Wolf Cukier\altaffilmark{1,2},Ravi kumar Kopparapu\altaffilmark{2,8}, Stephen R. Kane\altaffilmark{3,8}, William Welsh\altaffilmark{4}, Eric Wolf\altaffilmark{5,8}, Veselin Kostov\altaffilmark{2,6}, Jacob Haqq-Misra\altaffilmark{7,8}}

\altaffiltext{1}{Scarsdale High School, NY}
\altaffiltext{2}{NASA Goddard Space Flight Center, 8800 Greenbelt Road, Mail Stop 699.0, Building 34, Greenbelt, MD 20771}
\altaffiltext{3}{Department of Earth Sciences, University of California, Riverside, CA 92521, USA}
\altaffiltext{4}{Department of Astronomy, San Diego State University, 5500 Campanile Drive, San Diego, CA 92182, USA}
\altaffiltext{5}{Laboratory for Atmospheric and Space Physics, Department of Atmospheric and Oceanic Sciences, University of Colorado Boulder, Boulder, CO 80309, USA}
\altaffiltext{6}{SETI Institute, 189 Bernardo Avenue, Suite 200, Mountain View, CA 94043, USA}
\altaffiltext{7}{Blue Marble Space Institute of Science, 1001 4th Ave, Suite 3201, Seattle, Washington 98154, USA}
\altaffiltext{8}{NASA Astrobiology Institute's Virtual Planetary Laboratory, P.O. Box 351580, Seattle, WA 98195, USA}

\begin{abstract}
We use a one-dimensional (1-D) cloud-free climate model to estimate habitable zone (HZ) boundaries for terrestrial planets of masses 
0.1 M$_{E}$ and 5 M$_{E}$ around circumbinary stars of various spectral type combinations. Specifically, we consider 
binary systems with host spectral types F-F, F-G, F-K, F-M, G-G, G-K, G-M, K-K, K-M and M-M. Scaling the 
background N$_{2}$ atmospheric pressure with the radius of the planet, we find that the inner edge of the HZ moves
inwards towards the star for 5 M$_{E}$ compared to 0.1 M$_{E}$ planets for all spectral types.  
This is because the water-vapor column depth is smaller for larger planets and higher temperatures 
are needed before water vapor completely dominates the outgoing longwave radiation. The outer edge of the HZ
changes little due to competing effects of the albedo and greenhouse effect. While these results are broadly 
consistent with the trend of single star HZ results for different mass planets, there are significant differences
between single star and binary star systems for the inner edge of the HZ. 
Interesting combinations of stellar pairs from our 1-D model results can be used to explore for in-depth 
climate studies with 3-D climate models. 
We identify a common HZ stellar flux domain for all circumbinary spectral types. 

\end{abstract}

\maketitle

\section{Introduction}
Binary stars are ubiquitous in the galaxy, with nearly half of all Sun-like stars residing in binary
(and higher multiple star) systems. Numerous studies in the past two decades have predicted that
planets can form and sustain long-term stability around binary stellar systems (Alexander 2012;
Paardekooper et al. 2012; Pierens \& Nelson 2007, Meschiari 2012a; Marzari et al. 2013; Liu 2013;
Mason et al. 2013; 2015; Georgakarakos et al. 2015). The {\it Kepler} mission has detected several exoplanets
in binary systems (Doyle et al.2011, Welsh
et al. 2012; Orosz et al. 2012a, 2012b; Kostov et al. 2013; Schwamb et al. 2013; Welsh et al. 2015).
 Despite a strong observational bias against the discovery of such planets,
at the time of writing there are six confirmed planets orbiting one member of a sub-20 AU binary
stellar system (i.e. circumprimary planets or S-type systems, Kley \& Haghighipour 2014) and 12
confirmed planets orbiting within 3 AU of both members of sub-AU binary star systems (circumbinary
planets or P-type systems, e.g. Welsh et al. 2015, Kostov et al. 2016a). Based on known
circumbinary systems, estimates suggest a $1-10\%$ occurrence rate of Neptune- to Jupiter-sized
planets (e.g. Armstrong et al. 2014, Welsh et al. 2015, Kostov et al. 2016). Given the proximity to
their host star, planets in binary systems experience the effects of two incident stellar fluxes. Almost
half of known circumbinary planets reside in the “habitable zone (HZ)” (Doyle et al.
2011; Orosz et al. 2012a; 2012b; Welsh et al. 2015; Kostov et al. 2016), as constrained by existing
estimates of the HZ for binaries (Kaltenegger \& Haghighipour 2013; Haghighipour \& Kaltenegger
2013; Eggl et al; 2012; 2013; Kane \& Hinkel 2013; Forgan 2016; Wang \& Cuntz 2019).

Our goal in this study is to estimate the HZs around circumbinary terrestrial planets using 1-D climate models. 
We describe the methodology in section 2, present the results of our analysis in section 3, and provide a discussion of their implications in section 4.

\section{Model and Methods}
We used the 1-D radiative-convective, cloud-free climate model from the Kasting group, which has been updated
recently (Kopparapu et al. 2013a, 2014). Details of the model are given in these papers and references there in.
In order to simulate the stellar flux incident on a circumbinary planet, we
follow the methodology of Kane \& Hinkel (2013), where the combined stellar energy distribution (SED) from both
the stars is used to estimate the equivalent effective temperature of a single energy source
that would produce the same energy flux. We then ran the model 1000 times, each time at a different point in the orbit, where the
stellar SED was combined. We performed these calculations for 10 cases: F-F, F-G, F-K, F-M, G-G, G-K, G-M, K-K, K-M 
and M-M. It is assumed that the planet is in a circular orbit around the stars, and the two stars are not orbiting 
each other. We used he BT-Settl grid of models\footnote{\url{http://perso.ens-lyon.fr/france.allard/}}
\citep{Allard2003, Allard2007} Two end-member planetary masses are considered: 0.1 and 5 M$_{E}$, to be consistent with previous studies
(Kopparapu et al. 2014; Wang \& Cuntz(2019)). We scale the background N$_{2}$ atmospheric pressure with the radius 
of the planet, which suggests that larger planets should have thicker atmospheres. The corresponding scaling is
given in Kopparapu et al. (2014).

We followed the methodology from Kasting et al. (1993) and Kopparapu et al. (2013a) to estimate the HZs.
The inner edge of the HZ is calculated by increasing the surface temperature of a fully saturated Earth model from 
220 K up to 2200 K. The effective solar flux $S_{eff}$, which is the value of solar constant required to maintain a 
given surface temperature, is calculated from the ratio between the net outgoing IR flux and the net incident 
solar flux, both evaluated at the top of the atmosphere. When $S_{eff}$ asymptotes to a constant value, that is
when the atmosphere is optically thick to the outgoing IR radiation, and the planet enters the runaway greenhouse
regime. This is considered the inner edge of the HZ. The total flux incident at the top of the atmosphere is taken 
to be the present solar constant at Earth's orbit 1360 W m$^{2}$. The outer edge of the HZ is calculated by fixing the
surface temperature of an Earth-like planet with 1 bar N$_{2}$ atmosphere, and the atmospheric CO$_{2}$ partial pressure
 was varied from 1 to 35 bar. Due to competing effects of the outoging IR and the incoming solar, $S_{eff}$
experiences a minimum as a function of CO$_{2}$ partial pressure. This minimum is where the `maximum' amount of 
warming can be achieved with CO$_{2}$, and this is the 'maximum greenhouse' limit for the outer edge of the HZ.

\label{sec2}

\section{Results}

\begin{figure}
\subfigure[]{
\label{ihz}
\includegraphics[angle=270,width=.50\textwidth]{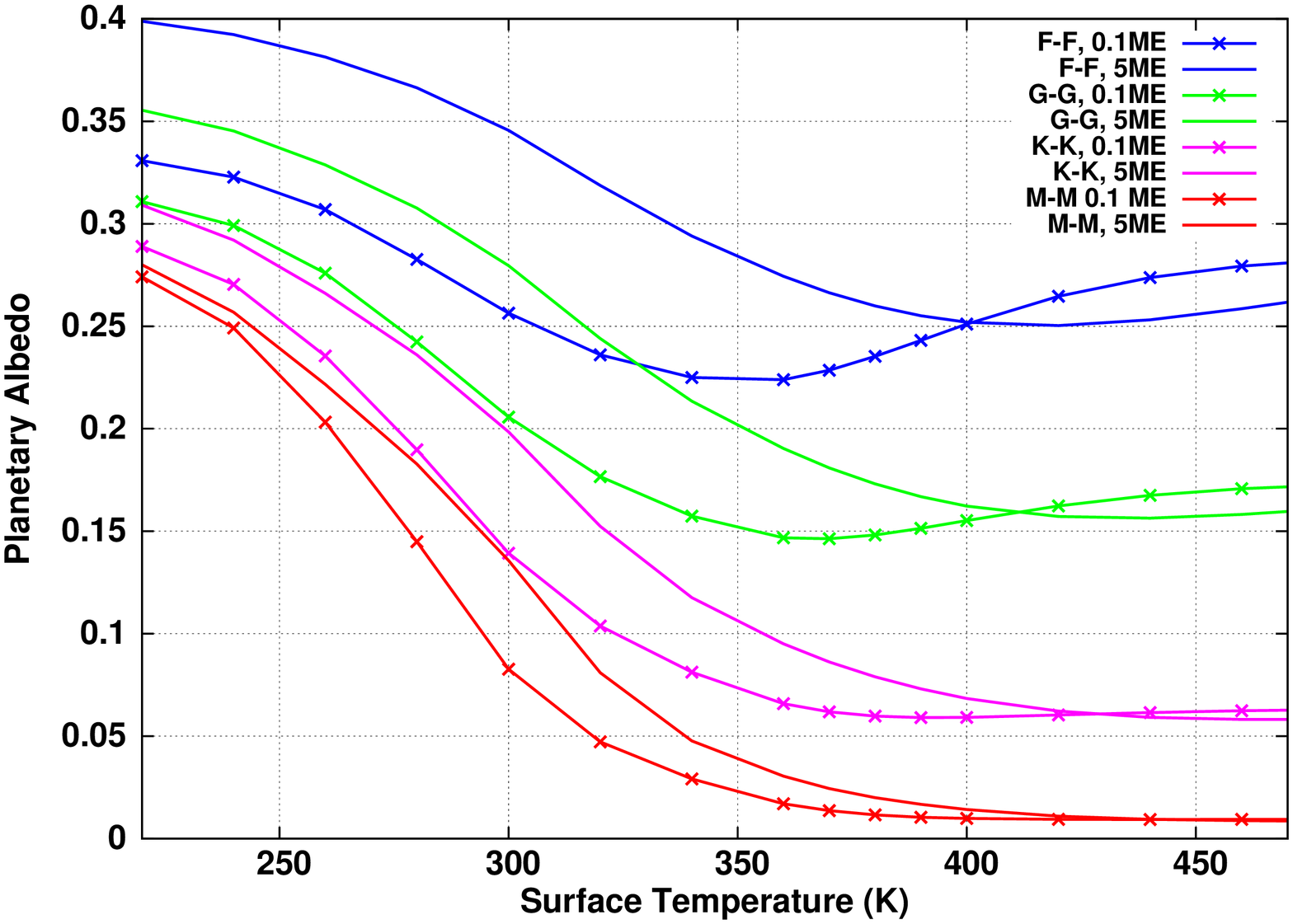}
}
\subfigure[]{
\label{ohz}
\includegraphics[angle=270,width=.50\textwidth]{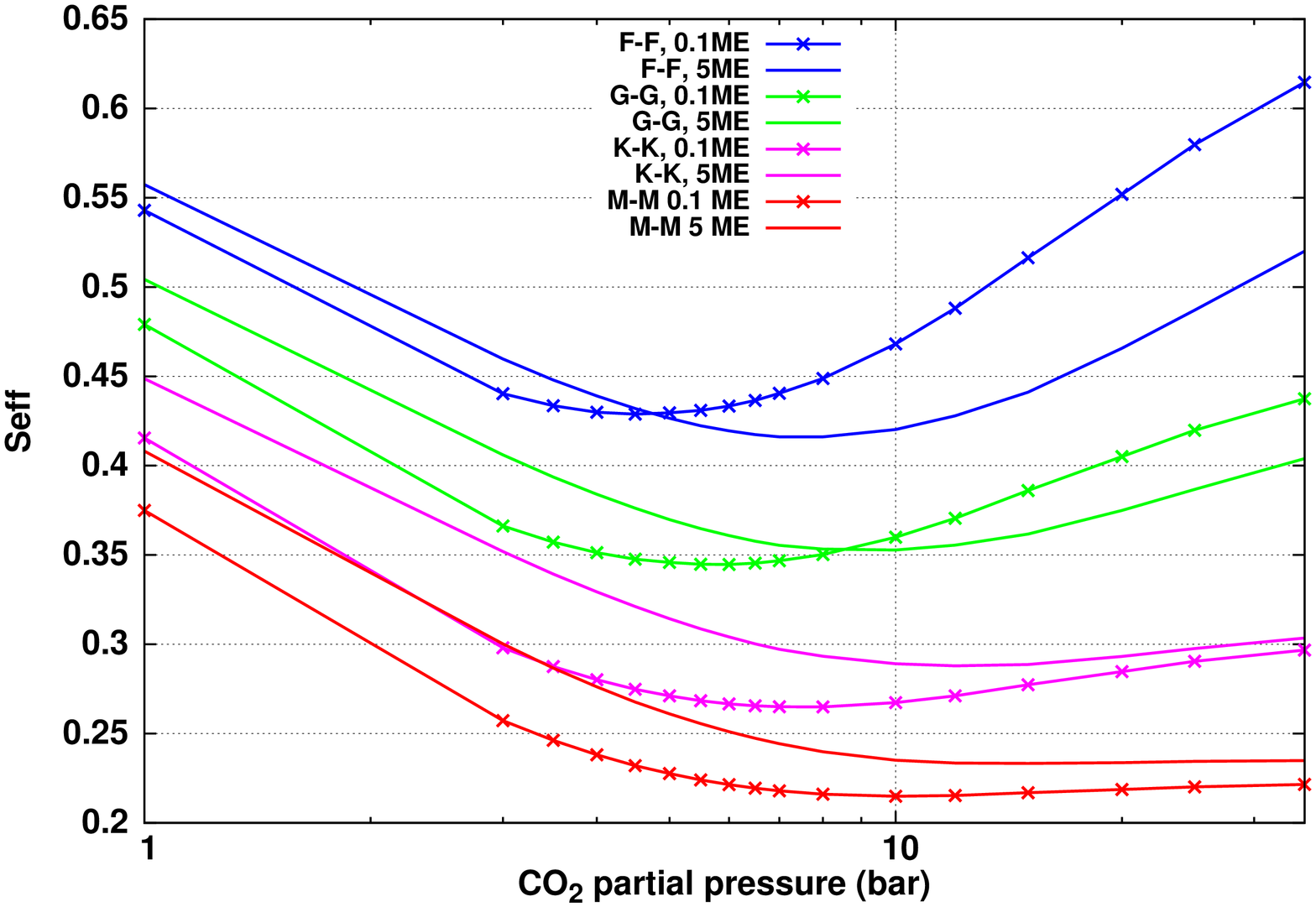}
}
\caption{(a): Top of the atmosphere planetary albedo as a function of surface temperature for a variety of spectral type configurations, and (b) effective stellar flux as a function of 
partial pressure of $\co2$,  for 0.1 M$_{E}$ (marker curve) and 5 M$_{E}$
(solid curve) planets around different host binary stars.}
%\label{hist}
\end{figure}

Figs. \ref{ihz} and \ref{ohz} show the results for inner and outer HZ, respectively, of 0.1 and 5 M$_{\oplus}$ planets
around binary stars of equal spectral types. Intermediate cases of mixed stellar spectral types 
(i.e, F-G, G-K, K-M etc.) fall within the regions of equal spectral types, and we do not show these results to 
maintain the clarity of our results shown in Figs. \ref{ihz} and \ref{ohz}. 
%Similar study for single stars was
%performed by Kopparapu et al. (2014), and by Wang \& Cuntz(2019) who used the results from Kopparapu et al. (2014),
%the climate focused results from a 1-D model were not included in any previous study. 

The planetary albedo shown in Fig.\ref{ihz} is higher if the host binary is comprised of hotter stars,
 and lower if the binary has cooler stars.  The reason is that the Rayleigh scattering cross section (inversely 
proportional to $\lambda^{4}$) is on average 
higher for planets around an F star, as the star's Wien peak is bluer (shorter wavelength) compared to cooler stars.
 Furthermore, $\h2o$ and $\co2$ have stronger absorption coefficients in the near infrared (NIR) than in the visible, 
so the amount of starlight absorbed by the planet's atmosphere increases as the radiation is redder (as is the case for 
an M-M binary). Both effects are more pronounced when the atmosphere is dense and full of greenhouse gaseous absorbers. 
Hence, for a planet around M-M binary, the planetary albedo is significantly lower due to minimal Rayleigh scattering
and high NIR absorption. Because we scaled the background $\n2$ pressure with planetary mass, 
as was done in \cite{Kopp2014}, there is a 
 higher amount of non-condensable gas on 5 M$_{E}$ than on a 0.1 M$_{E}$. This increases the Rayleigh scattering for
a 5 M$_{E}$ (solid curves in Fig. \ref{ihz}) compared to a 0.1 M$_{E}$ (marker curve). This effect is more 
pronounced around host binary stars that have peak radiation more blue shifted (for ex: F-F), 
as can be seen in Fig. \ref{ihz}. As the surface temperature increases, the planetary albedo decreases as a 
consequence of absorption of NIR solar radiation by $\h2o$. It then increases again and asymptotes 
at higher temperatures as Rayleigh scattering becomes dominant.

Fig. \ref{ohz} shows $S_{eff}$ as function of partial pressure of $\co2$, for our outer HZ calculations. The idea
here is to estimate the {\it maximum} amount of $\co2$ needed to maintain a surface temperature of 273K, and the 
corresponding energy balance needed to obtain this in the form of $S_{eff}$. As $\co2$ partial pressure increases,
$S_{eff}$ initially decreases because of the NIR absorption of $\co2$ greenhouse effect. However, beyond certain
amount of $\co2$ in the atmosphere, $\co2$ ice clouds start forming and gradually increase the Rayleigh 
scattering \citep{Kasting1993}. This effect is not pronounced at first because the greenhouse effect of $\co2$ dominates at lower amounts
of $\co2$. Remember that $S_{eff}$ is a ratio between the net outgoing IR flux and the net incident solar flux.
As the amount of $\co2$ increases outgoing IR flux asymptotically approaches a constant value as the atmosphere becomes 
optically thick at all IR wavelengths. However, the net incident solar flux decreases monotonically with increases in 
$\co2$ partial pressure as a result of increased Rayleigh scattering. Hence, $S_{eff}$ has a turnover, or a 
minimum at a corresponding partial pressure of $\co2$. This is the {\it maximum} amount of $\co2$ that can provide
a greenhouse warming, and hence the ``maximum greenhouse limit''.

Comparing 0.1 M$_{E}$ and 5 M$_{E}$ planets for the outer HZ case (Fig. \ref{ohz}), $S_{eff}$ is smaller for a 
lower mass planet, at low $\co2$ partial pressures. This is because the atmosphere of a 0.1 M$_{E}$ planet has a 
larger column depth, which increases the greenhouse effect and reduces the outgoing IR flux, decreasing $S_{eff}$. 
 At the same time, this larger column depth also increases
the planetary albedo at high $\co2$ partial pressures, {\it increasing} the $S_{eff}$ for a 0.1 M$_{E}$ planet.
A similar effect happens for the 5 M$_{E}$ mass planet, but here the atmosphere has a smaller column depth. Hence,
it increases the outgoing IR flux (effectively 'cooling' the planet) at low $\co2$ pressures, but at the same time
Rayleigh scattering is weak at high $\co2$ pressures.
The maximum greenhouse effect changes little because of these two competing effects, for a given host binary star
system.

\begin{figure}[!hbp]
%\centering
\subfigure[]{
\includegraphics[angle=360,width=0.50\textwidth]{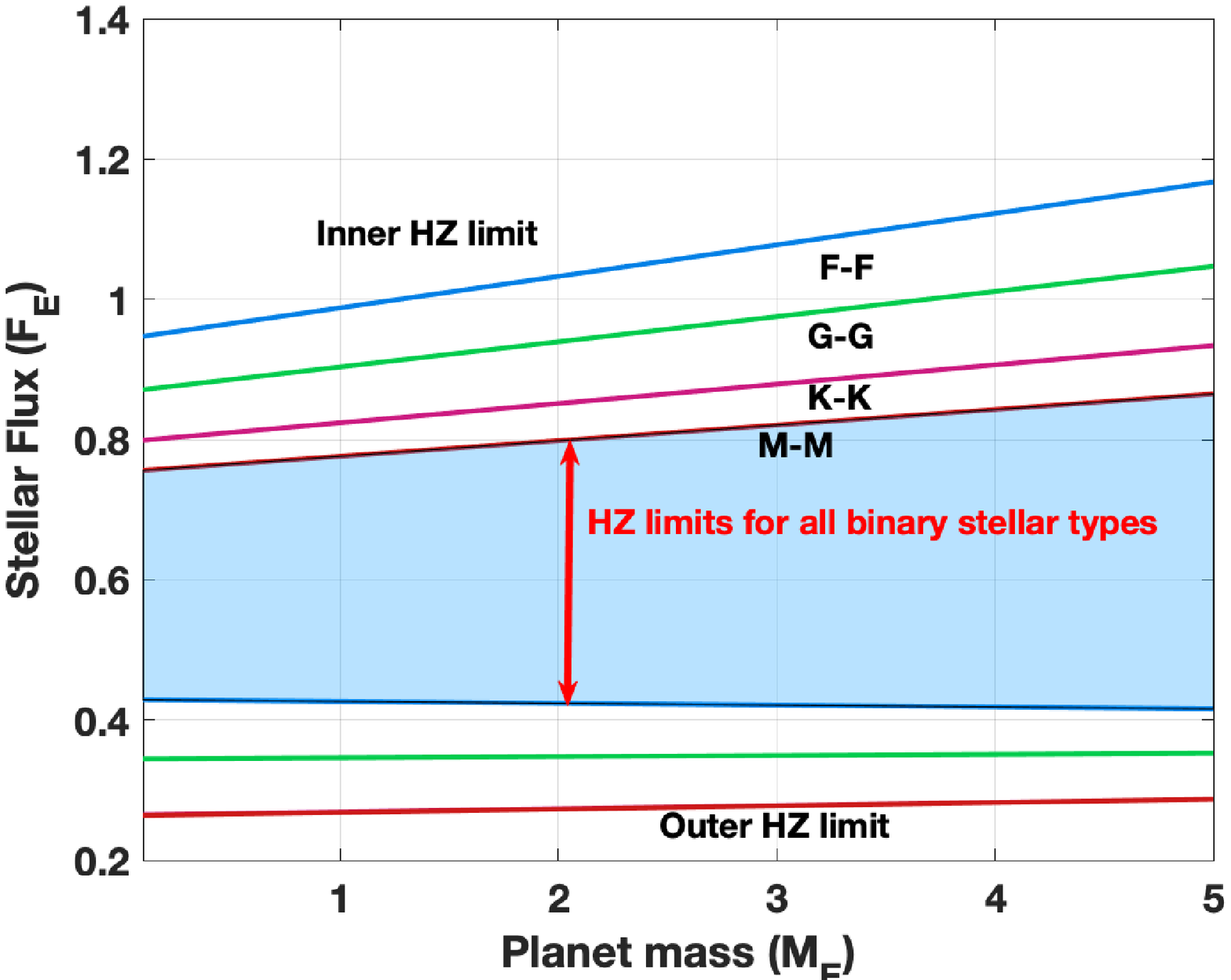}
}
\subfigure[]{
\includegraphics[angle=360,width=0.50\textwidth]{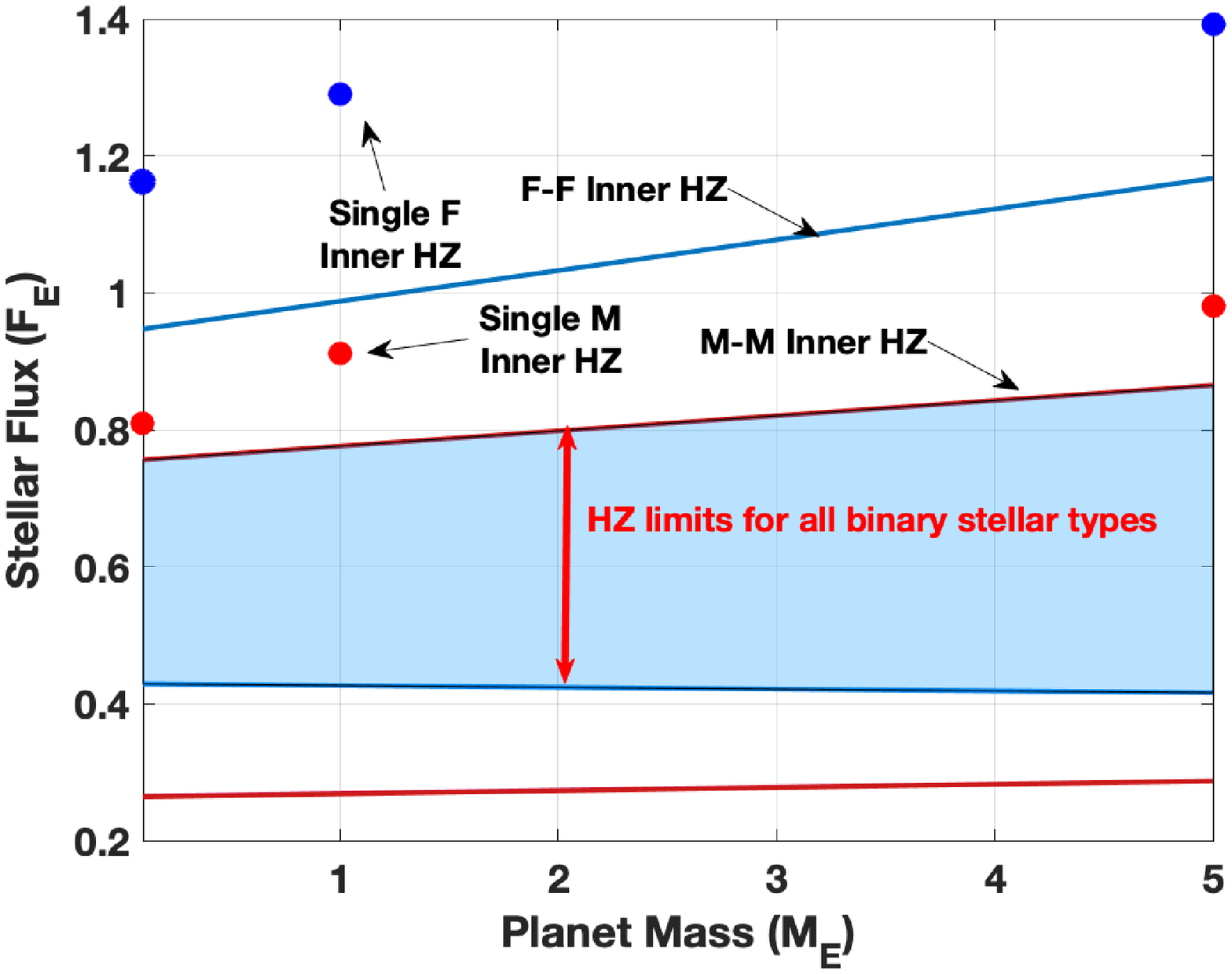}
}
\caption{HZ estimates from 0.1 to 5 Earth mass planets around all circumbinary stellar spectral types. The blue
shaded region is the width of the HZ that is common across all stellar binary spectral types. The left panel (a) shows HZ
limits for circumbinary star, while the right panel includes single star HZ stellar flux limits for comparison.
The inner HZ limit for binary stars occurs at lower stellar fluxes (farther from the stars) because of the
additional star contributing to the IR photons, and thus 'raising' the near-IR flux incident on the planet,
which increases the greenhouse warming. }
\label{hz_limits}
\end{figure}

Fig. \ref{hz_limits} show the summary of the results from Figs. \ref{ihz} \& \ref{ohz}, in terms of incident stellar
flux on the vertical axis, and planetary mass on the x-axis. Inner and outer HZs for F-F, G-G, K-K and M-M binary
star types are shown in panel (a) and (b). The HZ limits for mixed combination of stellar spectral type binaries 
are a subset of these
curves, and overlap within the ranges of the inner edge of the HZ for the F-F binary (top blue curve), and the outer
edge of the M-M binary (bottom red curve). The blue shaded region represents the HZ for {\it all} different binary
star and planetary mass scenarios. The inner edge of the HZ for a 5 M$_{E}$ occurs at relatively high stellar flux
compared to a 0.1 M$_{E}$ case, whereas the outer edge of the HZ changes very little. This result is consistent with
the single star HZ results from Kopparapu et al. (2014), where they found that the HZ for a massive planet is larger
compared to a smaller mass planet. 

For comparison, we have also included HZ estimates for single stars (specifically, F \& M spectral types) in 
Fig. \ref{hz_limits}. The inner HZ limit in stellar flux for single stars is larger compared to binary stars because
adding another star comparatively increases the number of photons available in the near-IR part of the combined
SED. This enhances the greenhouse warming in comparison with the Rayleigh scattering, and one need not have to 
'push' the planet as close to the star (higher stellar flux) as for a single star to drive the planet into 
runaway greenhouse regime.

\section{Discussion}
Our 1-D model results assume a circular orbit for the planet around our binary star configurations. However, several
exoplanets discovered around circumbinary stars have eccentric orbits. Previous studies
have included the effect of eccentricity on Earth-like planets in the HZs of circumbinary planets 
(Eggl et al. 2012, 2013; Kane \& Hinkel 2013; Mueller \& Haghighipour 2014) and even eccentric host binary 
stars (Kley \& Haghighipour 2014). These studies indicate that the orbit averaged flux incident on the planet
varies significantly, depending upon the eccentricity. The time varying flux could induce changes in the climate
of the planet, that could affect the habitability. While earlier studies have used the results of analytical or
1-D climate model results to account for the impact on habitability, few 3-D climate model results have been utilized
to study the global dynamics of the planet (Popp \& Eggl 2017). Future work using general circulation models (GCMs)
is currently in progress from some of the co-authors of this study.  

We have also compared our results with Eggl et al. (2013) and Kaltenegger \& Haghighipour (2013). Comparing the
data for a K-M stellar binary case (Haghighipour, private communication), we find that Our data produced similar results to the Kaltenegger and Haghighipour (2013). However, Our data, at lest for the inner edge of the HZ for a F-M binary,
appears to differ by $\sim 5\%$ to Eggl et. al. (2013). This could be significant, depending upon how large is the
planet. Kopparapu et al. (2014) found that the inner HZ for larger size planets move in by as much as $7\%$. The
difference between Eggl et al. (2013) and our study likely to have arisen in the method of calculation that they propose 
(Eggl 2018).

There are substantial physical effects that are being ignored in this simple 1-D model calculations (and any other
result that is based on a non-higher dimensional model). For example, several 3-D climate model results have
shown that slow-synchronously rotating planets develop thick substellar clouds due to weak Coriolis force. These
substellar clouds increase the planetary albedo, potentially maintaining habitable conditions at higher stellar 
fluxes which otherwise would not be possible (Yang et al. 2013a, 2014; Kopparapu et al. 2016; Way et al. 2016; 
Del Genio et al. 2018; Turbet et al. 2018). These general conclusions are more relevant at the inner edge of the 
HZ. At the outer edge, the models assume the regulation of the climate by the carbonate-silicate cycle. 
Recent calculations have suggested that
planets in the outer regions of the HZ may be less likely to maintain stable, warm climates, but instead may
 oscillate between
long, globally glaciated states and shorter periods of climatic warmth (Kadoya \& Tajika 2014, 2015;
Menou et al. 2015; Haqq-Misra et al. 2016). Such conditions, similar to ''Snowball Earth''
episodes experienced on Earth, would be detrimental to the development of complex land life.
CO$_{2}$ sequestration beneath water ice, owing to the high density of CO$_{2}$ ice (1.5 gcm$^{-3}$)
compared to H$_{2}$O ice (1 g cm$^{-3}$), could potentially reduce or eliminate the deglaciation episodes
\citep{Turbet2017, Ramirez2018}.  Limit cycles may also occur at higher stellar fluxes, near the inner 
edge of the HZ,  at low outgassing rates 
\citep{PM2017}.
Both of these effects, which impact the inner and the outer HZs, need to be considered to improve upon the HZ
estimates given in this work. 

\section{Conclusions}
\label{conclusions}

We have estimated the HZs of 0.1 and 5 Earth mass planets around circumbinary stars of various stellar spectral types 
using a 1-D radiative-convective climate model. We identify the width of the HZ that could be used for any spectral 
type of circumbinary star combination in circular orbit. We find that planetary alebdo plays a major role in 
determining the inner edge of the HZ, while the competing effects of NIR absorption and the Rayleigh scattering
of $\co2$ clouds at the outer edge make it less sensitive to the variations in these two parameters. However,
a more detailed study using 3-D climate model studies is needed, to properly consider the atmospheric circulation
and the corresponding effect on the habitability of terrestrial planets around circumbinary stars. The 
value in this 1-D study is identifying the interesting combinations of stellar pairs that would lead to 
interesting behavior in a GCM, as 1-D studies are more suited for exploring parameter space.

        \acknowledgements

     The authors gratefully acknowledge funding from NASA Habitable Worlds grant 80NSSC17K0741 for the support of this
study.
       NASA affiliates acknowledge supportfrom GSFC Sellers Exoplanet Environments Collaboration (SEEC), which is funded in part by the NASA PlanetaryScience Divisions Internal Scientist Funding Model. We thank reviewer Ramses Ramirez for
constructive comments which improved the manuscript.

\end{document}